\documentclass[11pt]{iopart}
\usepackage{iopams}  
\usepackage{epsfig,amsfonts,amsbsy}
\newcommand{\bra}{\left\langle}
\newcommand{\ket}{\right\rangle}
\newcommand{\pder}[2]{\frac{\partial #1}{\partial  #2}}

\newcommand{\der}[2]{\frac{d #1}{d  #2}}
\newcommand{\dert}[2]{\frac{d^2 #1}{d  #2^2}}
\newcommand{\bv}[1]{{\boldsymbol #1}}

\newcommand{\ep}{\epsilon}

\newcommand{\betac}{\beta_{\rm c}}
\newcommand{\sgn}{{\rm sgn}}

\newcommand{\pss}{p_{\rm ss}}
\newcommand{\pone}{p_{\rm one}}
\newcommand{\Hone}{H_{\rm one}}
\newcommand{\Zone}{Z_{\rm one}}

\newcommand{\ps}{P_{\rm ss}}

\begin{document}
%\rightline{ver. 14/12/30}

\title{Collective Dynamics from Stochastic Thermodynamics}
\author{Shin-ichi Sasa}
\ead{sasa@scphys.kyoto-u.ac.jp}
\address{
Department of Physics, Kyoto University, Kyoto 606-8502, Japan}
\date{\today}

%%%%%%%%%%%%%%%%%%
\begin{abstract}
From a viewpoint of stochastic thermodynamics, we derive equations 
that describe the collective dynamics near the order-disorder
transition in the globally coupled XY model and near the 
synchronization-desynchronization transition in the Kuramoto 
model. A new way of thinking is to interpret the deterministic
time evolution of a macroscopic variable as an external 
operation to a thermodynamic system. We then find that the 
irreversible work determines the equation for the collective
dynamics. When analyzing the Kuramoto model, we employ 
a generalized concept of irreversible work which originates from a
non-equilibrium identity associated with steady state thermodynamics.
\end{abstract}


%\maketitle

% stochastic thermodynamics 

\section{Introduction}

%% introduction %%

% stochastic thermodynamics % 

Since  the discovery of the fluctuation theorem 
\cite{Evans,Gallavotti,Kurchan}, 
non-equilibrium statistical mechanics, which aims at connecting microscopic  
mechanics with macroscopic properties under non-equilibrium conditions,
has been intensively studied. In particular, thermodynamic concepts  
such as heat, work, and entropy production are seriously re-considered 
so as to have a consistent thermodynamics framework for each realization 
of fluctuating quantities 
\cite{Sekimoto-book, Seifert-review}.
This framework 
has been referred to as {\it stochastic thermodynamics}. 
Owing to much effort, nowadays, it can be said that the foundation of 
stochastic thermodynamics has been established, and we should 
consider a next challenge based on the development of stochastic
thermodynamics.

%%  what we will do

In the present paper, we discuss collective dynamics in systems 
consisting of many elements. This topic
is of course one of important problems in non-equilibrium physics,
but one may wonder how this problem is related to stochastic 
thermodynamics. Here, the first purpose of this paper is to shed
light on the connection between collective dynamics and stochastic
thermodynamics. A key point is that 
the deterministic time evolution of a macroscopic variable 
is interpreted as an external operation 
to a thermodynamic system, and {\it the weak irreversible work 
is ascribed to a macroscopic friction force for the external system}. 
The last phrase is taken from p. 192 in Ref. \cite{Sekimoto-book}. 
The combination of the friction force and the thermodynamic force 
gives rise to the total force. When the total force is expressed
in terms of the order parameter, a differential equation of the
order parameter is determined.  

%% detail-1 

In section \ref{MFXY}, we shall  explain basic notions by analyzing 
the globally coupled XY model subjected to thermal noise. According
to equilibrium statistical mechanics, the order-disorder transition
point in this model is determined by a self-consistent equation 
for the order parameter characterizing the phase order. We then 
consider the time evolution of the order parameter near the  transition 
point. Since its characteristic time scale 
is much longer than other variables, we interpret the time
dependence of the parameter as a nearly quasi-static operation 
to the system. In the quasi-static limit, the so-called adiabatic 
theorem holds, which claims that the work
is equal to the free energy change. We find that this relation is
equivalent to the self-consistent equation for determining the 
transition point. Then, in nearly quasi-static processes, the 
irreversible work, which is defined as the difference
between the work and the free energy change, appears slightly.
Here, the irreversible work 
is characterized by a macroscopic friction constant. 
Since the irreversible work in nearly quasi-static processes is 
connected to fluctuations of irreversible work in the quasi-static 
processes, the friction constant is determined from the time correlation
of a thermodynamic force at the trivial state. By calculating the
friction constant, we obtain  a differential equation of 
the order parameter. 

%% detail-2 

This method is elegant but seems applicable to only thermodynamic 
systems. As another example of collective dynamics, 
in section \ref{Kuramoto-model}, 
we study the Kuramoto model which is the simplest model that 
describes the collective synchronization \cite{Kuramoto-book,RMP-Kuramoto}. 
However, there are 
neither  thermodynamics, equilibrium statistical mechanics, 
nor Hamiltonian in the Kuramoto model. The situation is rather
different from the globally coupled XY model. Nevertheless, when
we add a noise term to the Kuramoto model, the Langevin equation
for each element is  similar to that of the globally coupled 
XY model \cite{Sakaguchi}. 
Only difference is that there exists a non-equilibrium 
driving force in the Kuramoto model. Thus, from a viewpoint
of stochastic thermodynamics, the analysis of the noisy Kuramoto
model requires an extension of the irreversible  work and the 
fluctuation-dissipation relation. 
Here comes the steady state thermodynamics of Langevin equations
\cite{HatanoSasa}.
We already found  the generalization of the irreversible work  in 
transitions between two steady states by extending the Jarzynski
equality \cite{Jarzynski} to that valid in non-equilibrium systems. 
By using this extended equality, we  derive a formula of the friction
constant in terms of time correlation functions at the trivial
state. As a result, we obtain a differential equation 
of the order parameter near the transition point of the noisy 
Kuramoto model. Furthermore, we can take the noiseless limit of 
the equation. 

% merit 

It should be noted that the collective dynamics of globally 
coupled XY model and the Kuramoto model were studied by the
so-called bifurcation analysis using a center manifold theory
\cite{Kuramoto-book, Crawford, Chiba}. That is, in this paper,
we do not derive new equations of the order parameters, but
we present a simpler derivation method than previously known 
ones. In particular, if we already know the self-consistent
equation, we have only to calculate the friction constant
in terms of time correlation functions. The calculation is
quite elementary. Furthermore, by distinguishing ``static quantities''
such as the free energy and ``dynamic quantities'' such as 
the friction constant, we can gasp the problem in a clear 
manner. Therefore, we expect that the method will be applied to
systems for which the collective dynamics are not studied yet.  
In the last section, we argue such future  problems to be 
studied. Throughout this paper, the Boltzmann constant is set to unity,
and $\beta$ is always identified with $1/T$. 

%%%%%%%%%%%%%%%%%%%
%  MF-XY          %
%%%%%%%%%%%%%%%%%%%

\section{Globally Coupled XY model}\label{MFXY}

% MF-XY model 

\subsection{Equilibrium Statistical Mechanics}

% model and question

Let $\theta_i$ $(1 \le i \le N)$ be a phase variable of $i$-th element. 
We denote a collection of phases $(\theta_i)_{i=1}^N$ by $\bv{\theta}$
and define the Hamiltonian as
\begin{equation}
H(\bv{\theta})= - \frac{K}{N}\sum_{i , j} \cos(\theta_i-\theta_j).
\end{equation}
The canonical ensemble of the system is given by
\begin{equation}
p^{\rm can}(\bv{\theta})=\frac{1}{Z} e^{-\beta H(\bv{\theta})}.
\label{canonical}
\end{equation}
We want to derive the equilibrium value of 
the order parameter defined by 
\begin{equation}
r e^{i \varphi}\equiv \frac{1}{N} \sum_{j=1}^N  e^{i \theta_j }
\label{order}
\end{equation}
with $r \ge 0$.

% analysis

We first notice that the Hamiltonian is expressed as
\begin{equation}
H(\bv{\theta})= - Kr \sum_{i} \cos(\theta_i-\varphi).
\end{equation}
Although $r$ and $\varphi$  depend on $\bv{\theta}$,
we can assume that they take the equilibrium values
(with probability one) in the limit $N \to \infty $ 
owing to the law of large numbers. We fix $r$ and $\varphi$ 
to these values. We then write 
\begin{equation}
p^{\rm can}(\bv{\theta})=\prod_i \pone^{\rm can}(\theta_i;r,\varphi),
\end{equation}
where 
\begin{equation}
\pone^{\rm can}(\theta_i;r,\varphi)
=\frac{1}{\Zone(r)}e^{-\beta \Hone(\theta_i;r,\varphi)}
\end{equation}
with
\begin{equation}
\Hone(\theta;r,\varphi)=- Kr \cos(\theta-\varphi).
\label{one-H}
\end{equation}
$\Zone(r)$ is the normalization constant given by
\begin{equation}
\Zone(r)=\int_0^{2\pi }d\theta 
e^{\beta Kr \cos(\theta)}.
\label{z1}
\end{equation}
The equilibrium value of $r$ is then determined by
\begin{eqnarray}
r &=&  \lim_{N \to \infty} \frac{1}{N}
\sum_{j=1}^N \cos(\theta_j-\varphi) \nonumber \\
 &=& \int_0^{2\pi} d \theta 
\pone^{\rm can}(\theta;r,\varphi) \cos (\theta-\varphi)
\nonumber  \\
&=&  \frac{1}{\Zone(r)}
\int_0^{2\pi} d \theta e^{\beta Kr \cos(\theta)}\cos (\theta) 
\nonumber \\
&=& \frac{1}{\beta K}\pder{}{r} \log{\Zone(r)}.
\label{sce}
\end{eqnarray}
By expanding (\ref{z1}) in $r$, we obtain
\begin{equation}
\log \Zone(r)=\log(2\pi)+\frac{1}{4}(\beta Kr)^2-\frac{1}{64}(\beta Kr)^4 
+O(r^6).
\label{z:expansion}
\end{equation}
The self-consistent equation (\ref{sce}) becomes
\begin{equation}
r=\frac{1}{2}(\beta Kr)-\frac{1}{16}(\beta Kr)^3+O(r^5).
\end{equation}
This indicates that the transition inverse temperature $\betac$ 
for fixed $K$ is given by
\begin{equation}
\betac K=2. 
\end{equation}
Indeed, there are no other solutions than 
the trivial solution $r=0$ for $\beta < \betac$, 
while there is another solution for $\beta > \betac$.

\subsection{Collective Dynamics}

% model and problem

Next, we consider the collective dynamics of the order parameter.
We assume that the time evolution of $\theta_i$ is described by
the Langevin equation
\begin{eqnarray}
\der{\theta_i}{t}&=& -\pder{H}{\theta_i} +\xi_i \nonumber \\
     &=& -\frac{K}{N}\sum_{j=1}^N \sin (\theta_i-\theta_j) +\xi_i,
\label{model:1}
\end{eqnarray}
where $\xi_i$ is Gaussian-white noise satisfying 
\begin{equation}
\bra \xi_i(t) \xi_j(t') \ket =2T \delta_{ij}\delta(t-t').
\label{noise}
\end{equation}
The stationary probability density is the canonical 
distribution (\ref{canonical}).
The problem we want to solve is to obtain a differential
equation of the order parameter $re^{i \varphi}$. 

% explicit and general expression
% time dependent self-consistent equation

In order to set the problem explicitly, we assume 
the probability density at the initial time $t=0$ as
\begin{equation}
p_0(\bv{\theta})=\prod_i \pone^{\rm can}(\theta_i; r_0,\varphi_0)
\end{equation}
for a given $r_0$ and $\varphi_0$. The probability density
$p(\bv{\theta},t)$ at time $t$ is determined uniquely. Then,
in the limit $N \to \infty$, $r(t)$ and $\varphi(t)$ for each $t$ 
take definite values for almost all $\bv{\theta}$ with 
respect to $p(\bv{\theta},t)$.
We fix  functional forms of $r(t)$ and $\varphi(t)$ to those. 
Since we can rewrite (\ref{model:1}) as
\begin{equation}
\der{\theta_i}{t}= -K r \sin (\theta_i-\varphi) +\xi_i,
\label{one-Lan}
\end{equation}
the probability density at time $t$ is expressed as 
\begin{equation}
p(\bv{\theta},t)=\prod_i \pone(\theta_i,t),
\end{equation}
where $\pone$ is given by the solution of the Fokker-Planck equation
associated with (\ref{one-Lan}):
\begin{equation}
\pder{\pone(\theta,t)}{t}+\pder{}{\theta}
\left[-Kr \sin (\theta-\varphi)  \pone -T \pder{}{\theta} \pone \right]=0 
\label{FP}
\end{equation}
with the initial condition 
$\pone(\theta,0)=\pone^{\rm can}(\theta; r_0,\varphi_0)$.
Then, $r(t)$ and $\varphi(t)$ satisfy
\begin{equation}
r(t)e^{i \varphi(t)}= \int_0^{2 \pi} d\theta \pone(\theta,t) e^{i \theta},
\label{sce-dyn}
\end{equation}
which is regarded as a self-consistent equation for $r(t)$ and $\varphi(t)$.

% remark on the symmtery

Here, we make  a symmetry consideration. Suppose that 
$\varphi(t)=\varphi_0$. Then, we can derive the solution
as $\pone(\theta,t)=\tilde \pone(\theta-\varphi_0,t)$.
This means that $\varphi(t)=\varphi_0$
is a solution of the self-consistent equation. Below, we set
$\varphi(t)=\varphi_0=0$. 

% focus on the transition 

Now, we focus on the collective dynamics near the transition 
point. Explicitly, we set $\beta K=2+\epsilon $ with $|\epsilon| \ll 1$
for fixed $K$.  
We then expect that the slow dynamics of $r(t)$ are
characterized by a scaling form 
\begin{equation}
r(t)=\eta^b \bar r(\eta t),
\label{scaling:r}
\end{equation}
where $\eta \to 0$ and $t \to \infty$ with $\eta t=\tau$ fixed;
and $\bar r$ is a function whose functional form is independent of $\eta$. 
We also expect that $\eta$ is related to  $\epsilon$ as 
\begin{equation}
\eta =|\ep|^a.
\label{scaling:eta}
\end{equation}  
The question is to derive an equation for $\bar r$
and to determine the values of  $a$ and $b$. 

% quick review 
% strategy 

Mathematically, we have only to analyze $\pone$ near the 
transition point. One can apply  a center manifold theory
to this system. (See section 5.7 in Ref. \cite{Kuramoto-book}.)
Instead, we consider the problem from a viewpoint of 
stochastic thermodynamics.
Hereafter, $\bra \ \ket$ represents the expectation with 
respect to this initial distribution and the noise sequence 
$\xi$.  We also denote the expectation of $A(\theta)$ with
respect to $\pone(\theta,t)$ by $\bra A \ket_t$, and  $\bra 
\ \ket_{r}^{\rm can} $ represents the expectation of the
canonical ensemble with the Hamiltonian $\Hone(\theta;r)$.

\subsection{Stochastic Thermodynamics}

% how stochastic thermodynamics appear

We study the Langevin equation (\ref{one-Lan}), 
where $r$ is given as a function of time. 
We interpret the time dependent parameter as a control by 
an external system, without any feedback from the system. 
Concretely, the force $\Phi$ done by 
the external system is  defined as 
\begin{equation}
\Phi(\theta;r) \equiv \pder{\Hone(\theta;r)}{r}.
\label{Phi-def}
\end{equation}
By using (\ref{one-H}), we express $r(t)$ 
determined in (\ref{sce-dyn}) as
\begin{equation}
r(t)=-\frac{1}{K} \bra \Phi(r) \ket_t.
\label{trans-form}
\end{equation}
According to equilibrium statistical mechanics, we have
\begin{equation}
\bra \Phi(r) \ket_r^{\rm can} = \pder{F(r)}{r},
\end{equation}
where $F(r)$ is the free energy defined by
$F(r)= -T \log \Zone(r)$. The self-consistent
equation (\ref{sce}) is equivalent 
to $\bra \Phi(r(t)) \ket_t =\bra \Phi(r(t)) \ket_{r(t)}^{\rm can}$. 
This is  valid only in the limit $t \to \infty$,
and in general cases there should be  the irreversible 
work defined by
\begin{equation}
W_{\rm irr}= \int_0^t ds \der{r}{s}
\left[\Phi(\theta(s);r(s))-\pder{F(r(s))}{r(s)}
\right].
\end{equation}
We then obtain 
\begin{eqnarray}
\der{\bra W_{\rm irr} \ket }{t} 
&=&    
\der{r}{t}
\left[\bra \Phi(r(t)) \ket_t-\pder{F(r(t))}{r(t)}\right] \nonumber \\
&=& 
\der{r}{t}
\left[-K r(t) -\pder{F(r(t))}{r(t)}\right]. 
\label{result-gen}
\end{eqnarray}
The problem now becomes to evaluate the irreversible work in the 
stochastic system. 
The important property here is that the time scale of $r$ is much 
longer than the relaxation time of the probability density for the
Langevin equation (\ref{one-Lan}) 
near the transition point. 
That is, the control is assumed to be 
performed  as a {\it nearly  quasi-static} 
process, which enables us to develop a perturbation theory. 
Furthermore, owing to the recent progress on the stochastic thermodynamics,
we have several identities associated with thermodynamic works.
By utilizing one of them, 
we can simplify the calculation of the irreversible work.

% expression

Concretely, we start with the Jarzynski equality \cite{Jarzynski}
\begin{equation}
\bra e^{- \beta W_{\rm irr} }\ket =1.
\label{J-eq}
\end{equation}
See \ref{app-1} as for the derivation of a generalized version 
of (\ref{J-eq}). 
By combining $e^{-x}  \ge 1-x$ with the identity (\ref{J-eq}), we derive
$\bra W_{\rm irr} \ket \ge 0$, which corresponds to the second law of 
thermodynamics. Furthermore, from the identity (\ref{J-eq}), 
in the nearly quasi-static 
regime $\eta \to 0$, we have 
\begin{equation}
\bra W_{\rm irr} \ket = \frac{\beta}{2} \bra W_{\rm irr}^2 \ket +O(\eta^2),
\label{FDT}
\end{equation}
which corresponds to the fluctuation-dissipation relation. 
By taking the derivative with respect to $t$, we obtain
\begin{equation}
\der{\bra W_{\rm irr} \ket}{t} 
= \beta \der{r}{t} \int_0^t ds \der{r}{s} B(t,s) +O(\eta^3)
\label{wirr-evol}
\end{equation}
with 
\begin{equation}
B(t,s)= \bra \left[ \Phi(\theta(t),r(t))  -\pder{F(r(t))}{r(t)}\right] 
\left[ \Phi(\theta(s),r(s))  -\pder{F(r(s))}{r(s)}\right] 
\ket. 
\end{equation}
The correlation time of $\Phi(\theta, r(s))$ is controlled by $T$. 
Since $dr(s)/ds \simeq O(\eta)$ is much smaller than $T$,
(\ref{wirr-evol}) becomes 
\begin{equation}
\der{\bra W_{\rm irr} \ket}{t} 
= \gamma(r(t))  \left( \der{r}{t}\right)^2 +O(\eta^3)
\label{dissW}
\end{equation}
with a friction constant
\begin{equation}
\gamma(r)= \beta \int_0^\infty ds 
\bra \left[ \Phi(\theta(s),r)  -\pder{F(r)}{r}\right] 
\left[ \Phi(\theta(0), r)  -\pder{F(r)}{r}\right] 
\ket_r^{\rm can}. 
\label{gamma-form}
\end{equation}
Essentially the same expression was obtained in Ref. \cite{SekimotoSasa}.
Here, the expectation is taken over samples in which
$\theta(0)$ is chosen obeying the canonical ensemble 
with $r$, and $\theta(t)$ is determined from the
stochastic time evolution with fixed $r$.  
By combining (\ref{dissW}) with (\ref{result-gen}),
we have 
\begin{equation}
\gamma(r(t)) \der{r}{t} = -K r(t) -\pder{F(r(t))}{r(t)} +O(\eta^2). 
\label{irr-1st:2}
\end{equation}
This determines the time evolution of $r(t)$ uniquely. 
By using the expansion  (\ref{z:expansion}), 
we rewrite (\ref{irr-1st:2}) as 
\begin{equation}
\fl
\gamma(r(t)) \der{r}{t}
=-K r(t) + K \left[\frac{1}{2}(\beta Kr(t))-\frac{1}{16}(\beta Kr(t))^3
+O(r(t)^5) \right] +O(\eta^2).
\end{equation}
Recalling $dr/dt =O(\eta)$ and $\beta K=2+\epsilon$, we find that 
the exponents in (\ref{scaling:r}) and (\ref{scaling:eta}) 
are given by $a=1$ and $b=1/2$. 
The equation for $\bar r(\tau)$ with $\tau=\eta t$ is
\begin{equation}
\gamma(0) \der{\bar r}{\tau}
=\sgn(\epsilon) \frac{K}{2} \bar r-\frac{K}{2}\bar r^3
\end{equation} 
in the limit $\eta \to 0$ and $t \to \infty$. The positivity of 
$\gamma$ ensures the stability of the trivial solution $\bar r=0$ 
for $\epsilon <0$ and the non-trivial solution $\bar r  =1$ for 
$\epsilon >0$, respectively. We also note that $\gamma >0$ implies
the monotonic increment of $W_{\rm irr}$ (see (\ref{dissW})), which 
is a stronger property than the second law of thermodynamics.

% estimation of $\gamma$

Finally, we calculate $\gamma(0)$.  From the definition of $\gamma$
in (\ref{gamma-form}), we have 
\begin{equation}
\gamma(0)= \beta K^2 \int_0^\infty ds 
\bra  \cos \theta(s) \cos \theta(0) \ket_{r=0}^{\rm can}. 
\label{gamma0}
\end{equation}
Let $C(t)$ be $\bra \cos \theta(t) \cos \theta(0) \ket$ for
the free Brownian motion $d \theta/dt=\xi$ which corresponds
to the case $r=0$ in the Langevin equation (\ref{one-Lan}).  
We then derive
\begin{eqnarray}
\der{C}{t} 
&=& 
- \bra  \sin \theta(t) \circ \xi(t) \cos \theta(0) \ket_{r=0}^{\rm can}
 \nonumber \\
&=& 
- T \bra  \cos \theta(t) \cos \theta(0) \ket_{r=0}^{\rm can}  \nonumber \\
&=& 
- T C,
\end{eqnarray}
where the symbol $''\circ''$ represents the multiplication 
in the sense of Strotonovich. 
Since $C(0)=1/2$, we obtain $C(t)=\exp(-T t)/2.$ The substitution
of this result into (\ref{gamma0}) yields
\begin{equation}
\gamma(0)= \frac{\beta^2 K^2}{2},
\label{gamma-result}
\end{equation}
which is evaluated to be $2$ at the transition point. In sum,
the differential equation for $\bar r$ is 
\begin{equation}
\der{\bar r}{\tau}
=\sgn(\epsilon) \frac{K}{4} \bar r-\frac{K}{4}\bar r^3.
\end{equation} 
We guess that there should be some references reporting this result,
say around 1970, but we do not find them. As far as we searched,
explicit calculation was presented in Ref. \cite{Pikovsky} 
using bifurcation analysis. Note however that the numerical
coefficient of the non-linear term in Eq. (7) of Ref. \cite{Pikovsky} 
is not correct.

\section{Kuramoto model}\label{Kuramoto-model}

\subsection{Model}

We study the Kuramoto model \cite{Kuramoto-book,RMP-Kuramoto}
\begin{equation}
\der{\theta_i}{t}=
\omega_i -\frac{K}{N}\sum_{j=1}^N \sin (\theta_i-\theta_j),
\end{equation}
where $K >0$ and $\omega_i$ is a time-independent stochastic
variable obeying the probability density  $g(\omega)$. 
We assume that $g(\omega)=g(-\omega)$, $g(\omega) \le g(0)$, 
and the second derivative of $g(\omega)$ at $\omega=0$ is positive.
Note that $g(\omega) \to 0$ in $\omega \to \infty$ because $\int d \omega 
g(\omega)=1$. The collective synchronization 
occurs when $K > 2/(\pi g(0))$. This result was obtained by the
analysis of the self-consistent equation for the order parameter
(\ref{order}), which corresponds to (\ref{sce}) in the globally coupled
XY model. After that, Kuramoto and Nishikawa attempted to derive the 
equation that describes the collective dynamics in  the Kuramoto
model \cite{Kuramoto-Nishikawa}. However, it turned out that 
the problem was hard to be solved. Especially, even in the linear
regime around the trivial state $(r=0)$, the analysis was far from
trivial, as 
pointed out in Refs. \cite{Strogatz91,Strogatz92}. 
As one remarkable result, 
Ott and Antonsen derived the differential equation for the 
collective dynamics by noting a special solution of the non-linear 
equation of the distribution \cite{Ott}. Note that this method 
relies on a special property of the model \cite{sp-oa1,sp-oa2} and 
that it cannot be 
applied to general cases. Quite recently, Chiba has derived the 
equation of the order parameter  by mathematically developing 
a center manifold 
theory with a resonance pole \cite{Chiba}. 

% noisy Kuramoto

Since the difficulty originates from the 
deterministic nature of the dynamics,  its noisy version 
\begin{equation}
\der{\theta_i}{t}= \omega_i -
\frac{K}{N}\sum_{j=1}^N \sin (\theta_i-\theta_j) +\xi_i
\label{model:2}
\end{equation}
has also been studied, where $\xi_i$ is Gaussian-white noise 
satisfying (\ref{noise}). This model was first proposed 
by Sakaguchi \cite{Sakaguchi}. 
The self-consistent equation of the order parameter 
in this model was analyzed  and the non-trivial solution 
corresponding to 
the synchronized state was derived \cite{Sakaguchi}. 
Then, based on the linear stability analysis of the self-consistent
solutions in the noisy Kuramoto model \cite{Strogatz91,Strogatz92}, 
bifurcation analysis 
was performed so as to  obtain a differential equation of the
order parameter 
near the transition point \cite{Crawford}. (See Ref. \cite{Strogatz00}
for a story related to the development.)

% purpose

In this section, we study the collective dynamics near 
the transition point for the noisy Kuramoto model
from a viewpoint of stochastic thermodynamics.
We then consider the noiseless limit  $T \to 0$. 

\subsection{Setup of the problem}

% one-body

We start with re-expressing (\ref{model:2}) by 
\begin{equation}
\der{\theta_i}{t}= \omega_i -K r \sin (\theta_i-\varphi) +\xi_i.
\label{one-Lan-K}
\end{equation}
We denote by $\pone^{\rm ss}(\theta_i;r,\varphi,\omega_i)$ 
the stationary probability density for the Langevin equation
(\ref{one-Lan-K}) with $(r,\varphi)$ fixed. 
We follow the analysis in the previous section step by step. 

% self-consistent dynamical equation

We assume the probability density at the initial time $t=0$ as
\begin{equation}
p_0(\bv{\theta})=\prod_i \pone^{\rm ss}(\theta_i; r_0,\varphi_0,\omega_i)
\end{equation}
for given $r_0$ and $\varphi_0$. The probability density
$p(\bv{\theta},t)$ at time $t$ is determined uniquely. Then,
for each $t$, in the limit $N \to \infty$,  $r(t)$ 
and $\varphi(t)$ take definite values for almost all 
$\bv{\theta}$ with respect to $p(\bv{\theta},t)$.
We fix functional forms of $r(t)$ and $\varphi(t)$. Then, 
the probability density at time $t$ is expressed as 
\begin{equation}
p(\bv{\theta},t)=\prod_i \pone(\theta_i,t;\omega_i),
\end{equation}
where $\pone$ is given by the solution of the Fokker-Planck equation
associated with the Langevin equation (\ref{one-Lan-K}):
\begin{equation}
\pder{\pone(\theta,t;\omega)}{t}+\pder{}{\theta}
\left[(\omega-Kr \sin (\theta-\varphi))  
\pone -T \pder{}{\theta} \pone \right]=0 
\label{FP-K}
\end{equation}
with the initial condition 
$\pone(\theta,0;\omega)=\pone^{\rm ss}(\theta; r_0,\varphi_0,\omega)$.
Then,  $r(t)$ and $\varphi(t)$ satisfy
\begin{equation}
r(t)e^{i \varphi(t)}= \int d\omega g(\omega) 
\int_0^{2 \pi} d\theta \pone(\theta,t;\omega) e^{i \theta},
\label{sce-dyn-K}
\end{equation}
which is regarded as a self-consistent equation for 
$r(t)$ and $\varphi(t)$.
Without loss of generality, we assume $\varphi_0=0$.
Since $\pone^{\rm ss}(-\theta,t;-\omega)=\pone^{\rm ss}(\theta,t;\omega)$,
we find from (\ref{FP-K})that 
$\pone(-\theta,t;-\omega)=\pone(\theta,t;\omega)$. Then, 
(\ref{sce-dyn-K}) leads to $\varphi(t)=0$.
Hereafter, $\bra \ \ket_\omega$ represents
the expectation over initial conditions and noise sequences
in the Langevin equation  (\ref{one-Lan-K}) 
with the frequency $\omega_i=\omega$.
We  denote the expectation of $A(\theta)$ with
respect to $\pone(\theta,t;\omega)$ by $\bra A \ket_{t,\omega}$, 
and $\bra \ \ket_{r,\omega}^{\rm ss} $ represents the expectation 
with respect to $\pone^{\rm ss}(\theta; r,\omega)$.

% focus on the transition 

Now, let $K_{\rm c}(T)$ be the transition point of the coupling
constant for the model with $T$ fixed. We characterize 
the distance from the transition point by
\begin{equation}
\ep \equiv  \frac{K-K_{\rm c}}{K_{\rm c}}.
\end{equation}
Then, in the asymptotic regime $|\epsilon| \ll 1$,  
we expect that $r$ evolves slowly and this behavior may
be characterized by a scaling form (\ref{scaling:r}) with 
(\ref{scaling:eta}). 
The question is to derive a differential equation for $\bar r$
together with determining the values of  $a$ and $b$.

\subsection{Useful Identity}

% original Hatano-Sasa 

Differently from the previous section,
thermodynamic concepts such as the irreversible work
are not established for transitions between 
non-equilibrium steady states. 
Indeed, the Jarzynski equality  (\ref{J-eq}) is not available for
the Langevin equation  (\ref{one-Lan-K}) due to the existence of the driving
force $\omega_i$. We thus need to consider 
an extension of the Jarzynski equality (\ref{J-eq}).
This was proposed by Hatano and Sasa \cite{HatanoSasa}.
By defining 
\begin{equation}
\phi(\theta;r,\omega)=-\log \pone^{\rm ss}(\theta;r,\omega),
\end{equation}
they derived 
\begin{equation}
\bra 
e^{- \int_0^t ds  \pder{\phi(r(s),\omega)}{r(s)} \der{r}{s}  }
\ket_{\omega}  =1.
\label{Hatano-Sasa}
\end{equation}
It should be noted that (\ref{Hatano-Sasa}) becomes the 
Jarzynski equality (\ref{J-eq}) when the stationary probability
density is the canonical one. The quantity
\begin{equation}
Y \equiv T \int_0^t ds  \pder{\phi(r(s),\omega)}{r(s)} \der{r}{s}
\end{equation}
is interpreted as a generalized irreversible work in a 
process starting from steady state. (See Ref. \cite{SasaJSTAT} for
a review of an extended framework of thermodynamics on the basis
of (\ref{Hatano-Sasa}).)
Next, since there is no Hamiltonian, we 
consider a different formulation from that using 
the thermodynamic force (\ref{Phi-def}).
Concretely, by defining 
\begin{equation}
A \equiv -K\cos \theta,
\end{equation}
and by recalling (\ref{order}), we have 
\begin{equation}
-K r(t) = \int d\omega g(\omega) \bra A \ket_{t, \omega}.
\label{r-A}
\end{equation} 
The first step of the analysis is to estimate 
$ \bra A \ket_{t,\omega} $ in the nearly quasi-static 
regime $\eta \to 0$.
Here, as  essentially the same identity as (\ref{Hatano-Sasa}),
we derive 
\begin{equation}
\bra A \ket^{\rm ss}_{r(t),\omega}=
\bra A e^{- \int_0^t ds 
\pder{\phi(r(s),\omega)}{r(s)} \der{r}{s}  }\ket_{\omega}.
\label{Hatano-Sasa-2}
\end{equation}
(See \ref{app-1} for the derivation.) In the limit $\eta \to 0$, 
this identity yields 
\begin{equation}
\bra A \ket^{\rm ss}_{r(t),\omega}=
\bra A \ket_{t,\omega}
-  \int_0^t ds  \der{r}{s} \bra A(\theta(t)) 
\pder{\phi(\theta(s);r(s),\omega)}{r(s)}  \ket_{\omega}
+O(\eta^2).
\label{HS-fdr}
\end{equation}
A similar relation was proposed by using the identity (\ref{Hatano-Sasa}) 
\cite{PJP}.
Furthermore, by noting the time scale separation, we rewrite it as
\begin{equation}
\Gamma(r(t), \omega)\der{r}{t}
=
\bra A \ket_{t,\omega} -  \bra A \ket^{\rm ss}_{r(t),\omega}+O(\eta^2)
\label{sce-order}
\end{equation}
with
\begin{equation}
\Gamma(r, \omega)
=\int_0^\infty ds  \bra A(\theta(s)) \pder{\phi(\theta(0); r,\omega)}{r}  
\ket^{\rm ss}_{r,\omega}.
\label{Gamma}
\end{equation}
This is a generalized fluctuation-dissipation relation claiming that 
the friction constant is expressed as the time correlation function.

% last step 

Finally, multiplying $g(\omega)$ with the both hand sides of (\ref{sce-order})
and integrating them in $\omega$,  we obtain
\begin{equation}
\gamma(r(t)) \der{r}{t}
=-Kr(t)  - G(r(t)) +O(\eta^2),
\label{sce-order-sum}
\end{equation}
where we have used (\ref{r-A}); and $\gamma(r)$ and $G(r)$ are
expressed as 
\begin{eqnarray}
\gamma(r) &=& \int d \omega g(\omega) \Gamma(r, \omega), 
\label{gamma} \\
G(r) &=& -K \int d\omega g(\omega) \bra \cos \theta \ket^{\rm ss}_{r,\omega}.
\label{Gr}
\end{eqnarray}
Since $G(r)=-G(-r)$ (see \ref{s-pd}),  $G(r)$ can be expanded as
\begin{equation}
G(r) =-a_1 r-a_3 r^3+O(r^5).
\label{gr}
\end{equation} 
The transition point $K_{\rm c}(T)$ is determined by 
\begin{equation}
K_{\rm c}=a_1|_{K=K_{\rm c}},
\label{kc-con}
\end{equation}
which is  obtained from the condition that the linear term 
in the right-hand side of (\ref{sce-order-sum}) becomes zero.
By substituting  $K=K_{\rm c}(1+\ep)$ and $r=\eta^{1/2} \bar r(\eta t)$ with
$\eta=|\ep|$ into (\ref{sce-order-sum}), we obtain
\begin{equation}
\gamma(0) \der{\bar r}{\tau}
=\sgn(\epsilon) K_{\rm c} \bar r +a_3 \bar r^3
\label{dynamics-K}
\end{equation} 
in the limit $\ep \to 0$ and $t \to \infty$,
where $\gamma(0)$ and $a_3$ are evaluated at $K=K_{\rm c}(T)$.

\subsection{Calculation of $a_1$, $a_3$, and $\gamma(0)$}

We expand $\pone^{\rm ss}(\theta;r,\omega)$ as 
\begin{equation}
\pone^{\rm ss}(\theta;r,\omega)
=\frac{1}{2\pi}+\sum_{n=1}^\infty q_{n}(\theta;\omega)r^n.
\label{p-expand}
\end{equation}
From (\ref{Gr})  and  (\ref{gr}), we have
\begin{eqnarray}
a_1 &=&  K\int d\omega g(\omega)
\int_0^{2\pi} d\theta \cos \theta q_1(\theta;\omega), \\
a_3 &=&  K\int d\omega g(\omega)
\int_0^{2\pi} d\theta \cos \theta q_3(\theta;\omega).
\end{eqnarray}
By using expressions of $q_1$ and $q_3$, which are given in
\ref{s-pd}, we obtain 
\begin{eqnarray}
a_1 &=& \frac{K^2}{2}\int d\omega g(\omega)
\frac{T}{\omega^2+T^2},  \label{a1}            \\
a_3 &=& -T \frac{K^4}{4}\int d \omega g(\omega) 
\frac{T^2-2\omega^2}{(\omega^2+T^2)^2(\omega^2+4T^2)}. 
\label{a3}
\end{eqnarray}
The coefficients  (\ref{a1}) and (\ref{a3}) in the expansion 
(\ref{gr})  were already calculated \cite{Sakaguchi},
but the last term in Eq. (25) of Ref. \cite{Sakaguchi}
involves an error. By recalling (\ref{kc-con}), we explicitly
derive the transition point $K_{\rm c}(T)$ as
\begin{equation}
1=\frac{K_{\rm c}(T) }{2} \int d\omega g(\omega)\frac{T}{\omega^2+T^2}.
\end{equation}

% friction constant 

Next, we calculate the friction constant $\gamma(0)$.
By using the expansion of the stationary probability density 
(\ref{p-expand}), we have 
$ \phi= \log 2\pi -2 \pi q_1 r+O(r^2)$. We then obtain
\begin{equation}
\pder{\phi}{r}
=- K  \frac{1 }{\omega^2 +T^2}
\left( T \cos\theta +\omega\sin\theta  \right)+O(r)
\label{phi}
\end{equation}
in small $r$. (See \ref{s-pd} for the expression of $q_1$.)
By combining (\ref{Gamma}) with (\ref{phi}), 
the frequency dependent friction constant 
$\Gamma(0,\omega)$ is expressed as 
\begin{equation}
\Gamma(0, \omega)
=
 K^2  \frac{1 }{\omega^2 +T^2}
\int_0^\infty dt \left[T C(t) +\omega D(t)  \right],
\label{Gamma-c}
\end{equation}
where $C(t) = 
 \bra \cos\theta(t) \cos\theta(0)  \ket^{\rm ss}_{r=0,\omega}$ and 
$ D(t) =  \bra \cos\theta(t) \sin\theta(0)  \ket^{\rm ss}_{r=0,\omega}$.
Here, as shown in  \ref{app-2},  we derive
\begin{eqnarray}
\int_0^\infty dt C(t) &=& \frac{T}{2(\omega^2+T^2)},     \label{C-int}  \\
\int_0^\infty dt D(t) &=& -\frac{\omega}{2(\omega^2+T^2)}.\label{D-int}  
\end{eqnarray}
By substituting (\ref{C-int}) and (\ref{D-int}) into (\ref{Gamma-c}),
we obtain
\begin{equation}
\Gamma(0, \omega)
=
K^2  \frac{T^2-\omega^2 }{2(\omega^2 +T^2)^2}.
\end{equation}
Therefore, the friction constant (\ref{gamma}) with $r=0$ becomes 
\begin{equation}
\gamma(0)=K^2  \int d\omega g(\omega) 
\frac{T^2-\omega^2 }{2(\omega^2 +T^2)^2}.
\label{Gamma-result}
\end{equation}
When $g(\omega)=\delta(\omega)$, the result becomes
(\ref{gamma-result}) obtained in the previous section.

\subsection{Noiseless limit}

We consider the noiseless limit $T \to 0$.  We first rewrite 
$a_1$ as 
\begin{equation}
a_1 = \frac{K^2}{2}\int d\omega g(T \omega) \frac{1}{\omega^2+1}.
\end{equation}
This immediately yields
\begin{equation}
\lim _{T \to 0} a_1= \frac{ \pi K^2 }{2}g(0),
\label{a1-final}
\end{equation}
which leads to $K_{\rm c}= 2/(\pi g(0))$.
Similarly, we obtain
\begin{eqnarray}
\lim _{T \to 0} a_3  &=&  \lim _{T \to 0} \frac{K^4}{4T^2} 
\int d \omega g(T\omega) 
\left[
\frac{\omega^2}{(\omega^2+1)^2}- \frac{1}{\omega^2+4}
\right] \nonumber \\
     &=& 
\frac{\pi K^4}{16} g''(0),
\label{a3-final}
\end{eqnarray}
where the double prime represents the second derivative. 
Next, we evaluate the noiseless limit of $\gamma(0)$.
The method used in the estimation of $a_1$ and $a_3$ is not
effective here. The heart of the calculation is to note 
an identity
\begin{equation}
\int d\omega \frac{T^2-\omega^2 }{(\omega^2 +T^2)^2}=0.
\end{equation}
By using it, we rewrite (\ref{Gamma-result}) as 
\begin{equation}
\gamma(0)=K^2  \int d\omega (g(\omega) -g(0))
\frac{T^2-\omega^2 }{2(\omega^2 +T^2)^2}.
\label{Gamma-result-2}
\end{equation}
We then obtain
\begin{eqnarray}
\lim_{T \to 0} \gamma(0)
&=& -K^2 \int d\omega (g(\omega) -g(0))\frac{1}{2\omega^2}
    \nonumber \\
&=& -K^2 \int d\omega  g'(\omega) \frac{1}{2\omega}.
\label{Gamma-final}
\end{eqnarray}
By substituting (\ref{a3-final}) and (\ref{Gamma-final}) into
(\ref{dynamics-K}), the equation of $\bar r$
becomes
\begin{equation}
\left[ -K_{\rm c}^2 \int d\omega  g'(\omega) \frac{1}{2\omega} \right]
\der{\bar r}{\tau}
=\sgn(\epsilon) K_{\rm c} \bar r +\frac{\pi K_{\rm c}^4}{16} g''(0) \bar r^3,
\label{dynamics-K-final}
\end{equation}
which coincides with Eq. (7.97) in Ref. \cite{Chiba}.
The right-hand side corresponds to the self-consistent equation
obtained by Kuramoto \cite{Kuramoto-book}.
Furthermore, we remark 
\begin{equation}
\frac{\lim_{T\to 0} a_3}{\lim_{T \to 0} \gamma(0)}
= \frac{g''(0)}{4 \pi g(0)^2}
\left(  \int_0^\infty d\omega g'(\omega) \frac{1}{\omega} \right)^{-1}.
\label{ratio}
\end{equation}
This expression 
corresponds to Eq. (138) in Ref. \cite{Crawford}. Although 
the numerical coefficient of the latter is different from (\ref{ratio}),
there is no contradiction between the two, because  $|\alpha|$ 
in Ref. \cite{Crawford} is equal to $r /2\pi$ in this paper.
(This unusual convention can be understood from Eq. (36) and Eq. (95)
in Ref. \cite{Crawford}.)

% example 

More explicitly,  we study a case that $g(\omega)$ is a  Cauchy 
distribution 
\begin{equation}
g(\omega)=\frac{\Delta }{\pi} \frac{1}{\omega^2+\Delta^2}.
\end{equation}
By substituting it into the formulas given in (\ref{a1-final}), 
(\ref{a3-final}), and  (\ref{Gamma-final}), we obtain
\begin{eqnarray}
\lim _{T \to 0} a_1 = \frac{K^2 }{2 \Delta},  \label{a1-ex}  \\
\lim _{T \to 0} a_3 = -\frac{K^4 }{8 \Delta^3},  \label{a3-ex}
\end{eqnarray}
and 
\begin{equation}
\lim_{T \to 0} \gamma(0)= \frac{K^2 }{ 2 \Delta^2}.
\end{equation}
Since $K_{\rm c}=2 \Delta$ (that comes from (\ref{a1-ex})),
(\ref{dynamics-K}) becomes
\begin{equation}
2 \der{\bar r}{\tau}=\sgn(\epsilon) K_{\rm c} \bar r - K_{\rm c}\bar r^3.
\label{dynamics-K-ex}
\end{equation} 
Both the decay rate and the growth rate below and 
above the transition point are 
$|K-K_{\rm c}|/2$ in the original time scale, which 
are equal to the results of the linear stability  analysis 
\cite{Strogatz91, Strogatz92}.  The stationary solution 
above the transition point is equal to the  result
by Kuramoto \cite{Kuramoto-book}. Finally, the order
parameter equation presented in Ref. \cite{Ott} becomes 
(\ref{dynamics-K-ex}) near the critical point.

\section{Concluding Remarks}

% summary

In this paper, we have studied collective dynamics 
from a viewpoint of stochastic 
thermodynamics. The most important achievement is that
we can obtain the order parameter equation 
 (\ref{dynamics-K-final}) quickly.
The key step in the derivation is to
utilize the fluctuation-dissipation relation (\ref{HS-fdr})
that is derived from the non-equilibrium identity 
(\ref{Hatano-Sasa-2}). Owing to this identity,
we have only to calculate time correlation functions
for a free Brownian particle driven on a ring, in 
addition to the previously known self-consistent 
equations \cite{Kuramoto-book, Sakaguchi}.

% order of limit % deterministic vs stochastic 

As is understood from the derivation method, the noiseless 
limit $T \to 0$
should be taken after the scaling limit $\ep \to 0$ and $t \to \infty$ 
is considered. When both $T$ are $\ep$ are finite, our theory
provides a good approximation for $\ep \ll T$. On the contrary,
the calculation method cannot be applied to the noiseless Kuramoto 
model. Nevertheless, one may expect that the behavior  for the case 
$\ep \ll T \ll 1$ is close to that for $\ep \ll 1$ and $T=0$.
This expectation is true for some cases, but not always 
valid. For example, it was pointed out in Ref. \cite{Strogatz92} 
that when $g(\omega)$ is zero expect for 
$[-\omega_0, \omega_0]$ with some positive $\omega_0$, 
the order parameter in the noiseless
Kuramoto model relaxes to the trivial state in a power law form 
for $K < K_{\rm c}$, which is not observed for the case $\ep \ll T \ll 1$. 
We need to develop a different formulation if
we want to understand the behavior of the noiseless
Kuramoto model correctly \cite{pre}.

% variation 

Although we focus on the simplest model of  coupled oscillators,
one can study more general cases such that  the interaction 
includes higher harmonics e.g. $\sin (\theta_i-\theta_j)
+h  \sin 2(\theta_i-\theta_j)$. See Ref. \cite{Daido} for
a self-consistent  equation, Ref. \cite{Crawford-Davies} 
for the analysis using the center manifold theory of 
the noisy case, and Ref. \cite{Chiba-Nishikawa} for the 
generalized center manifold theory for the noiseless case. 
According to Ref. \cite{Chiba-Nishikawa}, the early attempts 
\cite{Daido, Crawford-Davies} have some mistakes. 
See Ref. \cite{Pikovski2013} for a recent study.
It should be
noted that the symmetry property that leads to $G(r)=-G(-r)$ and 
$\varphi=\varphi_0$ is broken for the case $h \not = 0$. This
makes the calculation complicated. More importantly, 
it was shown that the value of the critical exponent $b$
changes discontinuously in the noiseless limit.
It would be a good problem to obtain a fresh view 
of this phenomenon by applying the method in this paper. 

% finite size fluctuation

So far, we have assumed that $N \to \infty$. In finite but 
large $N$ cases, we naturally expect that 
small Gaussian noise is  added to the 
equation for the order parameter. We want to theoretically
derive this stochastic equation. For example, one may start 
with  the exact stochastic equation for the distribution 
\begin{equation}
p(\theta, t)=\frac{1}{N} \sum_{j=1}^N \delta (\theta_j(t)-\theta), 
\end{equation}
which is referred to as Dean's equation \cite{Dean}. 
Writing the path-integral expression for the history of $p$, 
one may combine the WKB analysis with the techniques in this paper. 
It is a challenging problem to complete the formulation. 
See Refs. \cite{Pikovsky,Daido-fluc, Chow-letter} for arguments on finite
size fluctuations. 

% finite dimension cases 

Obviously, the exact determination of the differential equation
for the order parameter relies on the mean field nature of 
the model. When we attempt to study models in finite dimensions, 
further techniques will be necessary so 
as to derive the time evolution of a spatially modulated
order parameter. Then, a local stationary distribution 
for given spatial configurations of $r$ and $\varphi$ should 
be a reference state or an unperturbed state. 
Although it is a highly non-trivial
problem to derive the equation, we should start
this analysis seriously, because we have the simplest
derivation of the collective dynamics in the mean-field model.
The collective dynamics of  coupled oscillators 
defined on random networks and complex networks are also
worthwhile to be studied \cite{Ichinomiya,Network}.

% fluid dynamics 

Finally, we briefly mention a recent work in which 
the Navier-Stokes equation is derived from Hamiltonian particles systems
using a non-equilibrium identity \cite{sasa}. This derivation method is 
formally correct and the most compact in existing approaches.  
Simplifying calculation enables us to extract the essence of the 
derivation problem, and thus we can now carefully review  
previous studies by Mori \cite{Mori}, Mclennan \cite{Mclennan}, 
Zubarev \cite{Zubarev},
and Esposito and Marra \cite{Esposito}, which will be reported 
elsewhere. However, the method in Ref. \cite{sasa}
involves some mathematical 
assumptions such as convergences of time correlation
functions. Now, look into the Kuramoto model again. If we set $T=0$ in 
the integrand of (\ref{Gamma-result-2}), the friction constant 
$\gamma(0)$ diverges. Thus, the formal calculation does not make
a sense. Similarly, in the argument of the hydrodynamic equation, 
we should check the well-defined nature of the dissipation constants.
Maybe related to this issue, we point out that arbitrarily small 
noise is introduced even in mathematically deriving the Euler equation
\cite{Yau}. 

% ending 

Stochastic thermodynamics formalizes thermodynamic concepts 
of fluctuating quantities. It is obvious from this definition 
that the framework is useful for analyzing small machines such as 
molecular motors. In addition to such direct application,
universal formulas found in stochastic thermodynamics may be 
applied to several non-equilibrium dynamics. We hope that 
this paper will stimulate many researchers who work on various subjects.

% acknowledgements
\ack

The author thanks H. Chiba, Y. Kawamura, and H. Nakao for 
their guidance to studies on the Kuramoto model. 
He also thanks  Y. Kawamura again and K. Sekimoto for their
comments on the draft. 
The present study was supported by KAKENHI Nos. 25103002 and 26610115, 
and by the JSPS Core-to-Core program ``Non-equilibrium dynamics of 
soft-matter and information.''

\appendix

\section{Derivation of the Identity (\ref{Hatano-Sasa-2})}\label{app-1}

% Markov chain

We consider a time-dependent Markov chain on a finite set $X$,
where a time series $(x_n)_{n=0}^N$ with $x_n \in X$
is generated by  a transition matrix $T(x_n \to x_{n+1}; \alpha_n)$ 
with a time dependent parameter $\alpha_n$.  We denote by 
$\ps(x;\alpha)$ the stationary probability of the transition 
matrix $T(x \to y; \alpha)$. That is, $\ps(x;\alpha)$ satisfies
\begin{equation}
\sum_x T(x \to y;\alpha) \ps(x;\alpha)=\ps(y;\alpha).
\end{equation}
We then define the dual transition matrix $T^*(x \to y; \alpha)$ by
\begin{equation}
\ps(x;\alpha)T(x \to y; \alpha)=\ps(y;\alpha)T^*(y \to x; \alpha).
\end{equation}
It should be noted that $\sum_x T^*(y \to x; \alpha)=1$. We then 
have a trivial identity 
\begin{eqnarray}
&& T^*(x_1\to x_0;\alpha_0) 
\cdots T^*(x_N \to x_{N-1};\alpha_{N-1}) \nonumber \\
&=& 
\frac{\ps(x_0;\alpha_0)}{\ps(x_1;\alpha_0)}
T(x_0 \to x_1; \alpha_0) 
\frac{\ps(x_1;\alpha_1)}{\ps(x_2;\alpha_1)}
T(x_1 \to x_2; \alpha_1) 
\cdots
\end{eqnarray}
Here, by multiplying $A (x_N)\ps(x_N; \alpha_N)$ to the both hand sides
and taking the summation over histories $(x_n)_{n=0}^N$, we obtain
\begin{equation}
\bra A \ket^{\rm ss}=
\bra \frac{\ps(x_1;\alpha_1)}{\ps(x_1;\alpha_0)}\cdots
\frac{\ps(x_N; \alpha_N)}{\ps(x_N;\alpha_{N-1})} A (x_N) \ket,
\label{identity}
\end{equation}
where $\bra {\cal A} \ket $ for a trajectory dependent quantity 
${\cal A}$  represents 
\begin{equation}
\fl
\bra {\cal A} \ket
=\sum_{(x_n)_{n=0}^N} 
\ps(x_0;\alpha_0)T(x_0\to x_1;\alpha_0) \cdots T(x_{N-1}\to x_N;\alpha_{N-1})
{\cal A}[ (x_n)_{n=0}^N].
\end{equation}
Note that (\ref{identity}) is also valid for Markov chains 
on real numbers. 

% Langevin case 

Next, we study the following  Langevin equation for a phase variable $\theta$:
\begin{equation}
\der{\theta}{t}=f(\theta;\alpha)+\xi,
\end{equation}
where $f(\theta+2\pi;\alpha)=f(\theta;\alpha)$ 
and $\xi$ is Gaussian white noise
satisfying $\bra \xi(t) \xi(t') \ket=2 T\delta(t-t')$. We denote 
the stationary probability density by $\pss(\theta;\alpha)$. We 
discretize the Langevin equation with a time interval $\Delta t$.
Since this defines the Markov chain on real numbers, we have the 
identity (\ref{identity}). Then, by taking the limit 
$\Delta t \to 0$, we obtain  the identity (\ref{Hatano-Sasa-2}).
When we set ${\cal A}=1$, it becomes the identity (\ref{Hatano-Sasa}). 

\section{Stationary probability density}\label{s-pd}

The stationary distribution of (\ref{one-Lan-K}) with $(r,\varphi)$ fixed
is determined from 
\begin{equation}
(\omega-Kr \sin \theta) \pone^{\rm ss}(\theta;r,\omega)
-T \partial_\theta \pone^{\rm ss}(\theta;r,\omega)=J(r,\omega),
\label{st-con}
\end{equation}
where $J(r,\omega)$ is a constant independent of $\theta$.
By substituting (\ref{p-expand}) into (\ref{st-con}), we have
\begin{equation}
\omega q_{n}-T \partial_\theta q_n=K \sin\theta q_{n-1}+J_n
\end{equation}
with $q_0=1/(2\pi)$ and $J_n$ is a constant. We solve this
equation iteratively. Concretely, we calculate $q_1$ and $q_2$ as 
\begin{eqnarray}
q_1 &=&  \frac{K}{2\pi} \frac{1 }{\omega^2 +T^2}
\left( T \cos\theta +\omega\sin\theta  \right), \\
q_2 &=&  \frac{K^2}{2\pi} \frac{1 }{2(\omega^2 +T^2)(\omega^2 +4T^2)}
\left[  (2T^2-\omega^2) \cos2\theta +3\omega T\sin 2\theta  \right] .
\end{eqnarray}
Noting that $q_3$ is written as 
\begin{equation}
q_3= b_{33} \cos 3 \theta +c_{33}\sin 3 \theta 
    + b_{31}\cos \theta+c_{31} \sin \theta,
\end{equation}
we calculate only $b_{31}$ as 
\begin{equation}
b_{31}= 
 \frac{K^3}{2\pi} \frac{T(2\omega^2-T^2)}
{2(\omega^2 +T^2)^2(\omega^2 +4T^2) }.
\end{equation}
As is understood from these calculation, one can prove 
\begin{equation}
\int_0^{2\pi} d\theta \cos \theta q_n(\theta;\omega)  =0
\end{equation}
for even integer $n$. This leads to $G(-r)=-G(r)$ from (\ref{Gr}).

\section{Derivation of (\ref{C-int}) and (\ref{D-int})}\label{app-2}

We study the simple Langevin equation
\begin{equation}
\der{\theta}{t}=\omega +\xi,
\label{Leq}
\end{equation}
where $\xi$ is Gaussian noise satisfying
$\bra \xi(t) \xi(t') \ket =2T \delta(t-t')$. We shall calculate 
the time integration of the correlation functions 
$ C(t) =  \bra \cos\theta(t) \cos\theta(0) \ket$ and
$D(t) =  \bra \cos\theta(t) \sin\theta(0) \ket$.
We first consider the time derivative of the correlation functions.
By using the equation (\ref{Leq}),  we have
\begin{eqnarray}
\der{C}{t} &=&   \omega D(t)-T C(t), \label{c-der}\\
\der{D}{t} &=&  -\omega C(t)-T D(t), \label{d-der}
\end{eqnarray}
where we  used $D(t) =  - \bra \sin\theta(t) \cos\theta(0) \ket$
and $C(t) =    \bra \sin\theta(t) \sin\theta(0) \ket$. We note that
$C(0)=1/2$ and $D(0)=0$. From 
(\ref{c-der}) and (\ref{d-der}), we obtain
\begin{equation}
\dert{C}{t} = -2T \der{C}{t}  -(\omega^2 +T^2) C(t),
\label{c-dert}
\end{equation}
where  $ dC/dt|_{t=0}= -T /2$. The time integration of (\ref{c-dert})
over the interval $[0,\infty]$
 leads to 
\begin{equation}
\int_0^\infty dt C(t) = \frac{T}{2(\omega^2+T^2)}.  
\end{equation}
The time integration of (\ref{d-der}) yields 
\begin{equation}
\int_0^\infty dt D(t) = -\frac{\omega}{2(\omega^2+T^2)}.
\end{equation}

\end{document}